\documentstyle[12pt,epsf]{article}
\newcommand{\be}{\begin{equation}}
\newcommand{\ee}{\end{equation}}
\newcommand{\ba}{\begin{eqnarray}}
\newcommand{\ea}{\end{eqnarray}}
\newcommand{\baa}{\begin{eqnarray*}}
\newcommand{\eaa}{\end{eqnarray*}}
\newcommand{\lab}[1]{\label{#1}}

\newcommand{\dis}{\displaystyle}

\newcommand{\bhat}{\hat{\beta}}

\begin{document}
\title{Prediction of Peptide Conformation by the
Multicanonical Algorithm$^1$}

\author{Ulrich H.E. Hansmann$^2$ and Yuko Okamoto$^3$}
\footnotetext[1]{\it To appear in the Proceedings of the
Sixth Annual Workshop on Recent Developments in Computer Simulation
Studies in Condensed Matter Physics, 22--26 Feb. 1993, Athens,
Georgia.}
\footnotetext[2]{\it Department of Physics and Supercomputer Computations
Research Institute (SCRI),
 The Florida State University, Tallahassee, FL 32306, USA.}
\footnotetext[3]{\it Department of Physics,
 Nara Women's University,Nara 630, Japan.}
\def\today{\ }
\maketitle

\begin{abstract}
We test the effectiveness of the multicanonical algorithm for the
tertiary structure prediction of peptides and proteins. As a simple
example we study Met-enkephalin.
The lowest-energy
conformation obtained agrees with that determined by other
methods such as Monte Carlo simulated annealing. But unlike
to simulated annealing
the relationship to the canonical ensemble remains exactly controlled.
Thermodynamic quantities at various temperature can be calculated from
one run.
\end{abstract}

A protein or a peptide is a molecule that consists of a chain of $N$ amino
acid residues. There are 20 different amino acids known in nature. When $N$
is large one calls the molecule a protein, otherwise a peptide.
The prediction of tertiary structures of proteins, which determine their
biological function,  from their primary
sequences remains one of the long-standing unsolved problems
(for recent reviews, see, for example, Refs.\ \cite{Rev4}).
It is widely believed
that this structure corresponds to the global minimum in the energy. So
the problem amounts to finding the global minimum energy out of a huge
number of local minima separated by high tunneling barriers.
Within the presently
available computer resources, the traditional methods such as molecular
dynamics and Monte Carlo simulations at  relevant
temperatures tend to get trapped in  local minima.  One of the
methods which which seem to  alleviate this multiple-minima problem is
simulated
annealing.\cite{SA}
However, a disadvantage of simulated annealing is   that there is no
established protocol for annealing and
a certain number (which is not known {\it a priori}) of runs
are necessary to evaluate the performance.
Moreover, the relationship of the obtained conformations to the
equilibrium canonical ensemble at a fixed temperature remains unclear.\\

These problems may be overcome by the multicanonical algorithm which
was recently proposed by Berg {\it et al.}\cite{MU}
Originally developed to overcome the supercritical slowing down of
first-order phase transitions,\cite{MU1}  it has also been
tested for systems with conflicting constraints such as
spin glasses.\cite{SG1,SG2,SG3}  The latter systems suffer
from a similar multiple-minima problem and it was claimed that the
multicanonical algorithm outperforms simulated annealing in these
cases.\cite{SG2}

The idea of this method is based on
performing Monte Carlo simulations in a {\em multicanonical}
ensemble\cite{MU,TV} instead of the usual (canonical) Gibbs-ensemble.
In the canonical ensemble, configurations at an inverse temperature
 $\bhat \equiv 1/RT$ are weighted with  the Boltzmann factor
$
{\cal P}_B (E)\ =\ \exp \left( - \bhat E \right) .
$
The resulting probability distribution is given by
\be
P_{B}(E)\ ~\propto ~n(E) {\cal P}_{B}(E)~,
\lab{pb}
\ee
where $n(E)$ is the spectral density.
In the {\em multicanonical} ensemble,\cite{MU,TV} on the other hand,
the probability distribution is defined in such a way that a
configuration with any energy enters with equal probability:
\be
P_{mu} (E) ~\propto ~ n (E) {\cal P}_{mu} (E) = {\rm const}.
\lab{pd}
\ee
Then it follows that the multicanonical weight factor should have the form
\be
{\cal P}_{mu} (E) ~\propto ~n^{-1} (E)~. \lab{e3}
\ee
In order to define a explicit form of this weight factor, we
introduce two parameters $\alpha (E)$ and $\beta (E)$ as follows:\cite{MU}
\be
{\cal P}_{mu} (E)
= {\rm exp} \Bigl\{-(\bhat + {\beta}(E))E - {\alpha}(E)\Bigr\}.
\lab{mupa}
\ee
For any fixed $\beta (E)$ and $\alpha (E)$ this leads to the
canonical weight factor with the inverse temperature
$\beta = \bhat + \beta (E)$, hence the name ``multicanonical''.
For a numerical simulation  one needs estimators for the
multicanonical parameters $\beta (E)$ and $\alpha (E)$. The iterative
procedure by which one can get such estimators is described
elsewhere.\cite{HY}
Once the multicanonical parameters  are
determined, one multicanonical run is in principle enough
to calculate all thermodynamic quantities by
re-weighting.\cite{FS}
Since in the multicanonical ensemble all energies enter with equal
probability  a simulation may overcome the barriers between
local minima by connecting back to the high temperature states. In this way
the global minimum can be explored.\\

In the present work we apply the multicanonical algorithm to the
problem of protein folding, the tertiary structure prediction of
peptides and proteins.
The purpose of this work is primarily to test the effectiveness
of the algorithm. For this reason we have studied one of the simplest peptide,
Met-enkephalin.  The
lowest-energy conformation for the potential energy function ECEPP/2
\cite{EC1} is known\cite{Enk} and analyses with Monte Carlo
simulated
annealing with ECEPP/2 also exist.\cite{EnkO,RSA4}\\

\begin{figure}
\centerline{\epsfbox{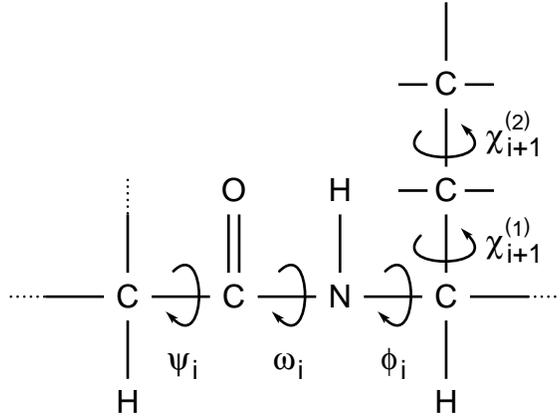}}
\caption{Definition of dihedral angles}
\label{fig1}
\end{figure}
Met-enkephalin has the amino-acid sequence Tyr-Gly-Gly-Phe-Met.
For our simulations the
backbone was terminated by a neutral NH$_2$-- ~group at the N-terminus
and a neutral~ --COOH group at the C-terminus as in the previous works of
Met-enkephalin.\cite{Enk}  The potential energy function
that we used is given by the sum of
the electrostatic term $E_{es}$, the van der Waals energy $E_{vdW}$, and
hydrogen-bond term $E_{hb}$ for all pairs of atoms in the peptide together with
the torsion term $E_{tors}$ for all torsion angles:
\begin{eqnarray}
E_{tot} & = & E_{es} + E_{vdW} + E_{hb} + E_{tors}\\
E_{es}  & = & \sum_{(i,j)} \frac{332q_i q_j}{\epsilon r_{ij}},\\
E_{vdW} & = & \sum_{(i,j)} \left( \frac{A_{ij}}{r^{12}_{ij}}
                                - \frac{B_{ij}}{r^6_{ij}} \right),\\
E_{hb}  & = & \sum_{(i,j)} \left( \frac{C_{ij}}{r^{12}_{ij}}
                                - \frac{D_{ij}}{r^{10}_{ij}} \right),\\
E_{tors}& = & \sum_l U_l \left( 1 \pm \cos (n_l \alpha_l ) \right).
\end{eqnarray}
$r_{ij}$ is the distance between the atoms $i$ and $j$, and $\alpha_l$ is
the torsion angle for the chemical bond $l$. For a definition of these
angles which represent the true degrees of freedom  see Fig.\ \ref{fig1}.
The parameters ($q_i,A_{ij},B_{ij},C_{ij},
D_{ij},U_l$ and $n_l$) for the energy function were adopted
from ECEPP/2,\cite{EC1}. The effect of surrounding atoms of water is
neglected and the dielectric constant $c$ is set equal to 2.
The computer code
KONF90,\cite{KONF}
was modified to accommodate the multicanonical method.  The peptide-bond
dihedral angles $\omega$ were fixed at the value 180$^\circ$
for simplicity,
which leaves 19 angles $\phi_i,\Psi_i$ and $\chi_i$ as independent variables.
\\

\begin{figure}
\vspace{5.5cm}
\includegraphics{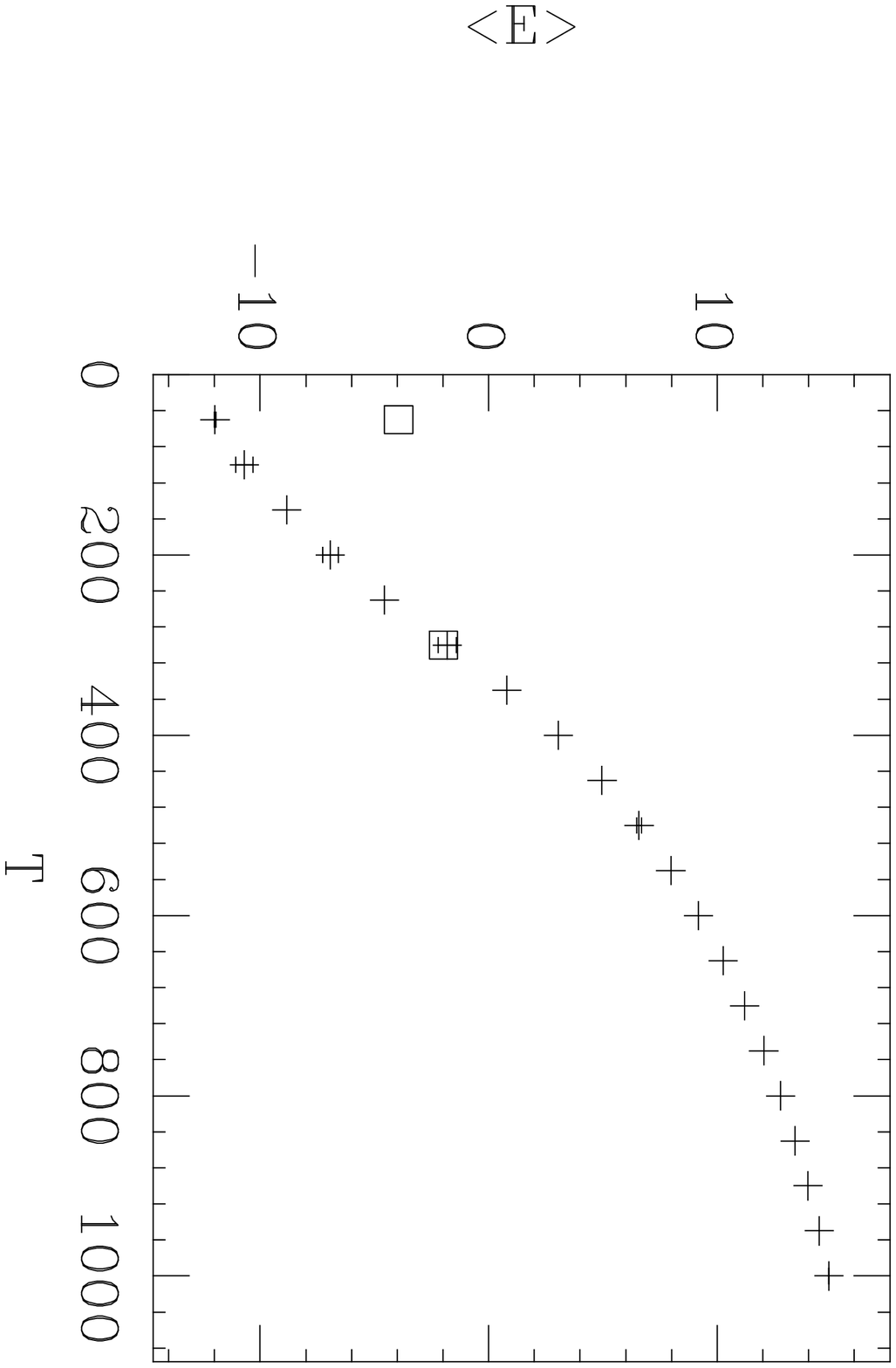}
\includegraphics{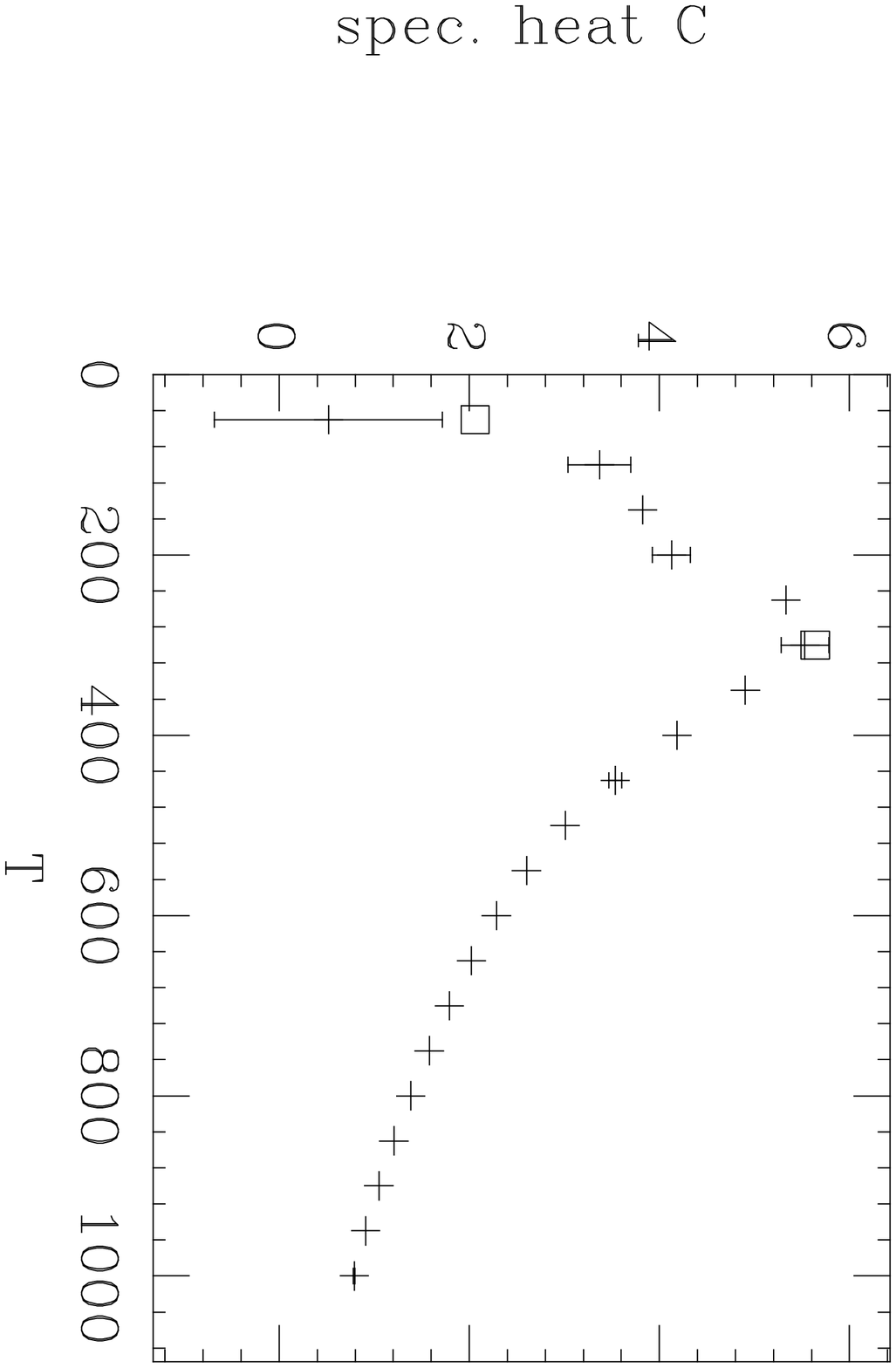}
\caption {Average energy $<E>$ and specific heat $C$ of Met-enkephalin as a
function of
temperature evaluated by multicanonical algorithms.  The results of canonical
simulations at fixed temperatures (50 K and 300 K) are also
plotted ($\Box$).}
\end{figure}
In Fig.~2a we show the average energy,  obtained by our method in a run
with $10^5$ sweeps, as a function of temperature.
The value $\approx -12$
kcal/mol at $T=50$ K is very close to the global-minimum energy obtained
by other methods.\cite{Enk,EnkO,RSA4}
In Fig.~2b we likewise present the \lq\lq specific heat" (per residue),
which is defined by
\be
C = \beta^2 ~\dis{\frac {<E^2>-<E>^2}{5}}~. \lab{e7}
\ee
It has a peak around $T=300$ K, which indicates that this temperature
is important for peptide folding.

During the production run the system reached the global-energy minimum
region  six times.  The lowest-energy conformation
within each visit is listed in Table~I together with the global-minimum
energy conformation (Conformation A in Table~I) obtained by simulated
annealing.\cite{EnkO}
Conformations 1--6 are the results at Monte Carlo steps
20128, 39521, 44462, 65412, 89413, and 95143.  Hence, the system reached
the lowest-energy region in every 5000 to 20000 Monte Carlo steps.  The
energies are almost all equal, and the lowest-energy value in the present
work ($-12.1$ kcal/mol) is slightly less than the previous result
($-11.9$ kcal/mol) by simulated annealing.\cite{EnkO}  Most of the dihedral
angles of the six conformations also agree with the corresponding ones
of Conformation A within $\approx 5^{\circ}$~.\\

{\small
\begin{table}
\caption{Table~I.  Energy and dihedral angles of the lowest-energy
conformations of Met-enkephalin obtained by multicanonical runs.
Conformation A is the
lowest-energy conformation obtained by Monte Carlo simulated annealing
(taken from Ref.~13). }
\begin{center}
\vspace{2ex}
\begin{tabular}{|c|r|r|r|r|r|r|r|} \hline
 Conformation  &   A~~ &  1~~  &  2~~  &  3~~  &   4~~ &    5~~ &  6~~\\ \hline
 E [ kcal/mol ]&$-11.9$&$-11.9$&$-12.0$&$-12.0$&$-12.1$&$-12.0$ &$-11.9$\\
\hline
 $\phi_1   $     &  98   & 90    &  91   & 90    & 97    & 96   & 98 \\
 $\psi_1  $      & 154   & 153   & 152   & 154   & 151   & 153  & 156 \\
 $\phi_2 $       &$-161$ &$-160$ &$-157$ &$-161$ &$-158$ &$-161$&$-163$\\
 $\psi_2      $  & 69    & 72    & 64    & 71    & 71    & 68   & 65 \\
 $\phi_3     $   & 65    & 64    & 66    &  63   & 64    & 64   & 66 \\
 $\psi_3    $    &$-93$  &$-95$  &$-92 $ &$-95 $ &$-94$  &$-89$ &$-92$\\
 $\phi_4   $     &$-85$  &$-82$  &$-80 $ &$-77 $ &$-83$  &$-85$ &$-80$\\
 $\psi_4  $      &$-27$  &$-26$  &$-29 $ &$-32 $ &$-30$  &$-31$ &$-29$\\
 $\phi_5 $       &$-83$  &$-81$  &$-82 $ &$-78 $ &$-80$  &$-82$ &$-86$\\
 $\psi_5$        & 142   & 142   & 138   & 137   & 145   & 151  & 147\\
 $\chi^1_1    $  &$-179$ & 179   &$-177$ & 179   & 179   &$-178$&$-176$\\
 $\chi^2_1   $   &$-112$ &$-110$ &$-117$ &$-109$ &$-111$ &$-115$&$-114$\\
 $\chi^3_1  $    & 149   & 144   & 146   & 143   & 149   & 145  & 142\\
 $\chi^1_4  $    & 180   &$-176$ & 178   & 177   & 180   &$-178$& 180\\
 $\chi^2_4  $    &  73   & 79    &  81   & 86    &  79   & 78   & 78\\
 $\chi^1_5  $    &$-65 $ &$-64 $ &$-67 $ &$-67 $ &$-66 $ &$-67$ &$-66$\\
 $\chi^2_5 $     & 180   &$-179$ & 180   & 180   &$-176$ & 180 & 176\\
 $\chi^3_5$      & 179   & 178   & 179   &$-179$ &$-179$ &$-178$&$-178$\\
 $\chi^4_5$      &$-55 $ &$-66$  &$-59 $ &$-62 $ &$-61$  &$-60$&$-57$\\
\hline
\end{tabular}
\end{center}
\end{table}
}
We have applied the recently developed multicanonical
algorithm to the problem of predicting the peptide conformation.
This method avoids getting trapped in a local minimum of energy function
by connecting
back to high temperature states and enhances in this way the probability
to find the global minimum.  We have demonstrated the effectiveness of the
algorithm by reproducing the lowest-energy conformation of Met-enkephalin.
Furthermore, the multicanonical algorithm can
yield various thermodynamic quantities as a function of temperature
from only one production run.\\
\
\\

\noindent
{\Large \bf Acknowledgements}:\\
Our simulations were
performed on the SCRI cluster of fast RISC workstations. This work is
supported, in part, by the Department of Energy, contract DE-AC03-76SF00515,
DE-FG05-87ER40319, DE-FC05-85ER250000 and by the
Deutsche Forschungsgemeinschaft under contract \hbox{H180411-1}. Y.O
likes to thank SCRI for the kind hospitality extended to him during
a visit.


\end{document}